\documentclass[conference]{IEEEtran}
\IEEEoverridecommandlockouts
\usepackage{cite}
\usepackage{amsmath,amssymb,amsfonts}
\usepackage{algorithmic}
\usepackage{graphicx}
\usepackage{textcomp}
\usepackage{url}
\usepackage{booktabs}
\usepackage[table,xcdraw]{xcolor}
\usepackage{colortbl}
\usepackage{listings}
\usepackage{multicol}
\usepackage{subfiles}

\definecolor{codegreen}{rgb}{0,0.6,0}
\definecolor{codegray}{rgb}{0.5,0.5,0.5}
\definecolor{codepurple}{rgb}{0.58,0,0.82}
\definecolor{backcolour}{rgb}{0.95,0.95,0.92}
\definecolor{Gray}{gray}{0.9}

\lstdefinestyle{mystyle}{
  backgroundcolor=\color{backcolour}, commentstyle=\color{codegreen},
  keywordstyle=\color{magenta},
  numberstyle=\tiny\color{codegray},
  stringstyle=\color{codepurple},
  basicstyle=\ttfamily\footnotesize,
  breakatwhitespace=false,         
  breaklines=true,                 
  captionpos=b,                    
  keepspaces=true,                 
  numbers=left,                    
  numbersep=5pt,                  
  showspaces=false,                
  showstringspaces=false,
  showtabs=false,                  
  tabsize=2
}
\def\BibTeX{{\rm B\kern-.05em{\sc i\kern-.025em b}\kern-.08em
    T\kern-.1667em\lower.7ex\hbox{E}\kern-.125emX}}
\begin{document}

\title{Just Dork and Crawl: Measuring Illegal Online Gambling Defacement in Indonesian Websites
}

\author{\IEEEauthorblockN{Luqman Muhammad Zagi}
\IEEEauthorblockA{
\textit{Institut Teknologi Bandung/ Radboud University}\\
Bandung, Indonesia / Nijmegen, The Netherlands \\
luqman.zagi@[itb.ac.id, ru.nl]}
\and
\IEEEauthorblockN{Girindro Pringgo Digdo}
\IEEEauthorblockA{
\textit{Cyber Army Indonesia}\\
Bandung, Indonesia \\
pringgo@cyberarmy.id}
\and
\IEEEauthorblockN{Wervyan Shalannanda}
\IEEEauthorblockA{
\textit{Institut Teknologi Bandung}\\
Bandung, Indonesia \\
wervyan@itb.ac.id}
}

\maketitle

\begin{abstract}
This study investigates the defacement of Indonesian websites by actors promoting illegal online gambling. Using a lightweight methodology that combines keyword-driven dorking with systematic crawling, we identified 453 defaced webpages within one month. Although dorking alone yielded a false positive rate of approximately 20.3\%, the integration of crawling and keyword-counting enabled reliable differentiation between true and false positives. Our measurements revealed diverse defacement behaviors, including repeat defacements (150 cases), fixed instances (129), keyword modifications (55), and redirections or hidden URL injections. In total, 8,837 unique third-party URLs spanning 5,930 domains were captured, with a small subset recurring across multiple sites. Website responses were inconsistent, with an average reaction time of 75.3 hours. These findings demonstrate that simple, reproducible techniques can provide meaningful insights into the scale, persistence, and dynamics of defacement, highlighting the importance of continuous measurement for strengthening defenses against online gambling activities.
\end{abstract}


\section{Introduction}

In recent years, government agencies and educational institutions in Indonesia have increasingly been targeted by cybercriminals who embed illegal gambling content into legitimate websites\cite{anupriyaIndonesianGovernmentWebsite2025}. Preliminary observations indicate that more than 2,500 websites were compromised within a four-month period by actors aiming to promote illegal online gambling platforms\cite{darkradarLaporanInsidenDefacement2025}. Such compromises can lead to significant reputation harm, erosion of public trust, and potential legal consequences for the affected entities\cite{holtExploratoryAnalysisCharacteristics2022}.

Globally, 
compromised websites has been systematically exploited to promote unregulated gambling, phishing, or scam-related content~\cite{yangCasinoRoyaleDeep2019}. Prior studies have addressed gambling site detection using URL and metadata analysis ~\cite{monevaRepeatVictimizationWebsite2022} as well as defacement detection using visual and HTML-based features~\cite{nguyenDetectingWebsiteDefacement2021}. However, many of these approaches depend on machine learning techniques that require extensive labeled datasets, significant computational resources, and continuous retraining\cite{nguyenDetectingWebsiteDefacement2021, vinayagamWebDefacementIdentification2024}, which make them difficult to deploy for real-time, large-scale monitoring.

In the Indonesian context, Illegal Online Gambling (IOG) defacement often extends beyond simple visual modifications. Operators typically inject external URLs~\cite{nursenoDetectingHiddenIllegal2024, darkradarLaporanInsidenDefacement2025}, hidden redirects~\cite{wuCloakingRedirectionPreliminary2005}, cloaking scripts~\cite{ibrahimLESSONLEARNPENANGANAN2024}, or keyword stuffing to manipulate search visibility and drive traffic to gambling platforms~\cite{nursenoDetectingHiddenIllegal2024,harahapDetectionOnlineGambling2024}. These techniques are often subtle and designed to persist undetected, which makes them harder to identify through conventional monitoring.

To address these challenges, we adopt a lightweight approach based on keyword-driven dorking and systematic crawling. This method enables large-scale, reproducible measurements without the need for resource-intensive infrastructure, making it suitable for continuous monitoring. Using this approach, we collected a dataset in May 2025 consisting of detected URLs and associated metadata, including timestamps, IP addresses, and keyword frequency counts. Our objective is to identify patterns of gambling-related defacement and evaluate the responsiveness of compromised websites. The findings of this study contribute to understanding the dynamics of IOG campaigns in Indonesia and provide practical insights for strengthening institutional defenses and incident response.

In summary, we make the following contributions
\begin{itemize}
    \item We present a lightweight and reproducible methodology for detecting defaced webpages, combining keyword-driven dorking with systematic crawling.
    \item We reveal that attackers employ diverse techniques, such as redirection, hidden URLs, and keyword manipulation, rather than solely relying on direct visual alterations of webpage content.
    \item We provide empirical measurements of website responses to defacement incidents, offering insights into remediation practices and response times.  
\end{itemize}

\section{Background and Related Works}
\subsection{Background}
\textbf{Website Defacement} is a form of cyber-dependent crime, involves unauthorized modification of website or web application content, altering their appearance and functionality~\cite{nguyenDetectingWebsiteDefacement2021, holtExploratoryAnalysisCharacteristics2022, monevaRepeatVictimizationWebsite2022}. 
Motivations for such attacks include amusement, patriotism, political or ideological agendas, revenge, and reputation building \cite{romagnaHacktivismWebsiteDefacement2017, holtExploratoryAnalysisCharacteristics2022, monevaRepeatVictimizationWebsite2022, burrussWebsiteDefacerClassification2022}. Research has shown that website defacement often recurs on the same target, with the same attacker responsible for multiple incidents, exhibiting a median of two attacks per affected website~\cite{monevaRepeatVictimizationWebsite2022}.

Common attack vectors fall into three primary categories. The first is \textit{vulnerability exploitation}, which targets weaknesses such as cross-site scripting (XSS) to inject malicious scripts or SQL injection to manipulate backend databases\cite{romagnaHacktivismWebsiteDefacement2017, vandeweijerHeterogeneityTrajectoriesCybercriminals2021}. The second is \textit{social engineering}, where techniques such as phishing and spoofing deceive administrators into revealing credentials or granting access\cite{vinayagamWebDefacementIdentification2024}. The third is \textit{technical compromise}, including the deployment of malware to gain persistence, DNS spoofing to redirect traffic to attacker-controlled servers, and the exploitation of unpatched or misconfigured systems to overwrite website content\cite{polonusWebsiteDefacedBecause2015, vandeweijerHeterogeneityTrajectoriesCybercriminals2021}.

\textbf{Gambling} in Indonesia is broadly prohibited under national law. The Criminal Code (Kitab Undang-Undang Hukum Pidana) explicitly criminalizes gambling under Article 303, punishing both organizers and facilitators of gambling activities, with possible imprisonment for those involved. Online gambling is subject to the same blanket prohibition. This is further stipulated in the Electronic Information and Transactions Law (Undang-Undang Informasi dan Transaksi Elektronik) Article 27 Paragraph 2, which prohibits any person from intentionally and without authorization distributing, transmitting, or making accessible electronic information and/or electronic documents containing gambling content\cite{indonesiaUndangundangUUNomor2008}.

\subsection{Related Works}\label{sec:relatedWorks}
Studies across Asia reveal the scale, complexity, and global connections of illegal online gambling. In South Korea, Min and Lee~\cite{minIllegalOnlineGambling2024} applied machine learning to textual and visual data to detect Absolute Illegal Online Gambling (AIOG) sites. Yang et al.\cite{yangCasinoRoyaleDeep2019} examined gambling promotion strategies and infrastructures, noting misuse of payments, outsourced services, and cloud storage. In Taiwan, Huang et al.\cite{huangStudyIllegalOnline2022} linked online gambling to organized crime and money laundering. In China, Gao et al.~\cite{gaoDemystifyingIllegalMobile2021} uncovered thousands of illicit apps and domains, revealing large-scale abuse of third-party services and over 140,000 suspicious servers. Collectively, these studies underscore the technological sophistication, transnational reach, and persistent challenges of combating illegal online gambling. Beyond Asia, research in Europe demonstrates the impact of regulataion on online gambling markets. Aonso-Diego et al.~\cite{aonso-diegoImpactSpanishGambling2025} showed that Spain's Royal Decree 958/2020, which restricting advertising, bonuses, and sponsorship, significantly reduced new gambling accounts, total money wagered, and marketing expenditures.

Website defacement detection is a longstanding cybersecurity challenge, as even short-lived compromises can disrupt services and damage credibility. Early techniques such as checksum comparison and DOM-tree analysis work for static pages but struggle with dynamic content and are costly at scale \cite{hoang2018website, albalawi2022website}. Machine learning approaches improved accuracy, with traditional models like Random Forests and Extra Trees achieving over 89–93\% accuracy and low false positive rates \cite{hoang2018website, kumar2023performance}. Deep learning further advanced detection, with Bi-LSTM models reaching up to 99\% accuracy \cite{ayunda2023comparative}, and multimodal approaches combining text and screenshots attaining 97.5\% accuracy with 1.5\% false positives \cite{nguyenDetectingWebsiteDefacement2021}. However, these methods often incur high computational cost and degrade under domain shift.
Hybrid systems have also emerged, such as WebTD, which integrates machine learning with blockchain verification to maintain over 98\% detection accuracy \cite{du2022fine}, and MeerKAT, which applies computer vision to webpage screenshots and achieves up to 98.8\% true positives with very low false positives on millions of samples \cite{borgolte2015meerkat}. Other adaptive methods, such as genetic programming, have been proposed to handle evolving environments \cite{medvet2007detection}.
\section{Methods}
Prior studies have employed machine learning techniques to detect website defacement §\ref{sec:relatedWorks}. However, such approaches are often complex and computationally intensive \cite{nguyenDetectingWebsiteDefacement2021}, requiring large annotated datasets, substantial training resources, and continuous model updates to remain effective. In this work, we adopt a simpler methodology based on keyword-driven search (dorking\footnote{The term Google dork was introduced by Johnny Long in 2002, originally referring to ``an inept or foolish person as revealed by Google'' \cite{longGoogleDorksArchived2002}.}) and systematic crawling. This approach is lightweight, scalable, and transparent, making it easier to reproduce and maintain while still effective for detecting widespread defacement incidents.

\begin{figure*}
    \centering
    \includegraphics[width=0.8\linewidth]{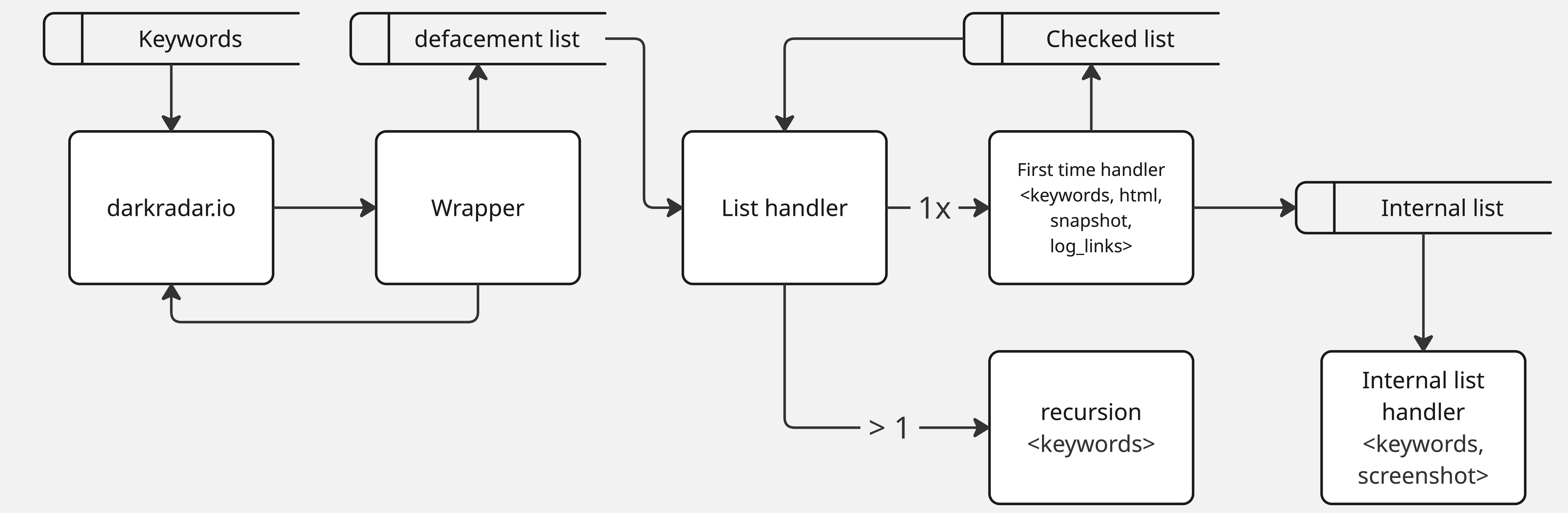}
    \caption{An overview of the data collection pipeline. The handler and recursion are based on DuckDuckGo's Tracker Radar Collector~\cite{Trackerradarcollector}}
    \label{fig:methods_pipline}
\end{figure*}

\subsection{Defaced Webpages List} \label{Sec:MethodsDefacedWebList}
To generate a list of defaced websites, we utilized DarkRadar\footnote{available on \url{https://www.darkradar.io}}, a tool developed by CyberArmy.id. Its core methodology employs Google Dorking to detect instances of website defacement. When predefined conditions are satisfied, the system issues alerts to notify users. Furthermore, DarkRadar preserves snapshots of compromised webpages by storing cached versions within its internal database.

We employed a set of keywords frequently found on defaced websites associated with illegal gambling activities \cite{harahapDetectionOnlineGambling2024, nursenoDetectingHiddenIllegal2024}. The selected keywords included ``togel'', ``toto'', ``judi'', ``slot'', ``gacor'', ``bandar'', ``maxwin'', ``zeus'', and ``bet''. The crawling process targeted domains under \texttt{.id}, with data extraction conducted at \textbf{two-hour} intervals.

\subsection{Defaced Website and Internal URL Data Collection}
All data collection in this study was conducted using DuckDuckGo’s Tracker Radar Collector (TRC) \cite{Trackerradarcollector}, an open-source and flexible web crawler framework that has been widely used in prior measurement studies. To meet the objectives of this work, we developed several custom collectors. 
The collectors implemented in this study included \texttt{keywords}, \texttt{HTML}, \texttt{snapshot}, and \texttt{iog\_link}.

The data collection pipeline was orchestrated by two main handlers. The \texttt{list handler} operated on an hourly basis. It retrieved the latest defaced webpage list from DarkRadar and verified whether each URL had appeared previously by comparing it against a \textit{checked list}. If a URL appeared for the first time, TRC was executed with all of the implemented collectors. This process have outputs of the \texttt{list handler} were twofold: (i) an updated \textit{checked list}, which was used for deduplication in subsequent executions, and (ii) an \textit{internal list}. For URLs that had already been recorded, the \texttt{list handler} executed TRC solely with the keywords collector to continue monitoring keyword activity.

The \textit{internal list} consisted of hyperlinks located within the same parent structure as the suspected IOG keywords in the HTML pages. This list included only third-party URLs, defined as domains different from the original landing page. The \texttt{internal list handler} processed this list once per day. For each entry, TRC was executed with the keywords and screenshot collectors to capture keyword activity and visual representations of the webpages.

\subsection{Defaced Websites Check} \label{sec:methods_flag}
To determine whether a webpage was defaced, we first excluded addresses beginning with \texttt{chrome}, as these indicated unreachable pages and were assigned a value of zero. For the remaining pages, we computed the total frequency of IOG-related keywords. Cases with only a single keyword were treated as false positives. Defacement likelihood was further assessed using page titles and URLs: if either contained IOG-related keywords, the page was classified as highly likely to be defaced.

\textbf{Defacement flag}
The collected webpages were classified based on the persistence and behavior of defacement indicators. \textit{Repeat defacement} refers to cases where defacement was initially detected, disappeared, and later reappeared. \textit{Fixed} denotes webpages that showed no defacement by the end of the measurement period, regardless of earlier fluctuations or temporary patches. \textit{Defaced (fluctuating)} captures instances where keyword counts varied over time but never declined to zero, indicating repeated content modifications. \textit{Defaced (constant) }describes cases where keyword counts remained unchanged from first detection until the end of observation. Lastly, the \textit{not found} category represents false positives, where no IOG-related content was confirmed after filtering and manual verification.
\section{Website Defacement}

\subsection{Crawler Overview}
The system was operated from 27 May 2025 to 25 June 2025 in Jakarta using a virtual machine configured with eight CPU cores and 8 GB of RAM. During this period, the crawler identified 453 webpages potentially affected by defacement. Upon verification, 15 of these pages failed to load, with 8 URL displaying keywords indicative of IOG activity. In total, 346 webpages were confirmed as defaced within the observation window, corresponding to 147 unique websites (Tab.~\ref{tab:crawlerOverview}). The number of defaced webpages per site ranged from 1 to 33, with a mean of 2.35 and a median of 1. This contrasts with Bartoli et al. \cite{bartoliReactionTimeWeb2009}, who reported that only 27.5\% of websites experienced a single defacement event.

Breaking down the confirmed defacements by domain, we identified eight top-level domains out of the 13 Indonesian TLDs managed by PANDI~\cite{pandiDomainID2025}. The largest share of affected webpages was observed in \texttt{.ac.id} (117 pages), \texttt{.go.id} (115 pages), and \texttt{.co.id} (46 pages). Defacements were also found in \texttt{.id}, \texttt{.or.id}, \texttt{.sch.id}, \texttt{.desa.id}, and \texttt{.web.id}, with a combined total of 68 pages. When grouped by unique websites, \texttt{.ac.id} accounted for 65 domains, \texttt{.go.id} for 39, and \texttt{.id} for 22, while the remaining TLDs each represented fewer than eight domains. This distribution highlights the concentration of defacements in academic and governmental domains, which may be more visible to attackers yet also more actively monitored.

\begin{table}[]
\caption{Crawler overview} \label{tab:crawlerOverview}
\centering
\begin{tabular}{@{}lr@{}}
\toprule
Pages found from dorking & 453     \\
\rowcolor[HTML]{EFEFEF} 
Failed to load & 15     \\
False positive          & 20.3\% \\
\rowcolor[HTML]{EFEFEF} 
Defacement pages       & 346     \\
Defacement websites         & 147     \\ 
\rowcolor[HTML]{EFEFEF} 
Total Top Level Domain & 8 \\
\bottomrule
\end{tabular}
\end{table}

\subsection{Keywords Distribution}

\begin{table}[]
\centering
\caption{Keywords occurrences from the defacement websites.} \label{tab:keywordsOccurrences}
\begin{tabular}{@{}lrr@{}}
\toprule
\multicolumn{1}{c}{Keywords} & \multicolumn{1}{c}{Websites} & \multicolumn{1}{c}{\begin{tabular}[c]{@{}c@{}}Max per pages\end{tabular}} \\ \midrule
slot   & 95.2\% & 715 \\
\rowcolor[HTML]{EFEFEF} 
gacor  & 89.1\% & 196 \\
toto   & 58.5\% & 227 \\
\rowcolor[HTML]{EFEFEF} 
judi   & 40.1\% & 103 \\
bandar & 42.9\% & 504 \\
\rowcolor[HTML]{EFEFEF} 
maxwin & 51.0\% & 141 \\
togel  & 48.3\% & 99  \\
\rowcolor[HTML]{EFEFEF} 
zeus   & 27.2\% & 37  \\
bet    & 90.5\% & 589 \\ \bottomrule
\end{tabular}
\end{table}

\begin{table}[]
\centering
\caption{Defacement flag found in webpages. The definition of the flag can be found on §\ref{sec:methods_flag}} \label{tab:flagOccurance}
\begin{tabular}{@{}lr@{}}
\toprule
\multicolumn{1}{c}{Flag} & \multicolumn{1}{c}{Webpages} \\ \midrule
Repeat Defacement        & 150                          \\ \rowcolor[HTML]{EFEFEF} 
Fixed                    & 129                          \\
Defaced (fluctuating)    & 55                           \\ \rowcolor[HTML]{EFEFEF} 
Defaced (constant)       & 12                           \\ \bottomrule
\end{tabular}
\end{table}

Although nine keywords were used for detection (listed in §\ref{Sec:MethodsDefacedWebList}), their occurrence levels varied considerably, as shown in Tab.\ref{tab:keywordsOccurrences}. The most prevalent keyword 
was ``slot'' (95.2\%), derived from slot machines, a popular gambling game typically featuring five reels that spin for several seconds\cite{harriganAddictiveGameplayWhat2010}. This was followed by ``bet'' (90.5\%), while the least prevalent was ``zeus'' (27.2\%). 
The term ``Zeus'' is commonly used in slot themes because of the widespread familiarity of mythological figures, which creates an instant connection and engagement for players\cite{netlingoExploringGlobalMythologies}.
In terms of intensity, the highest single-page count was observed for ``slot'', appearing 715 times on https://jim.unindra.ac.id/oai/. In contrast, ``zeus'' appeared on the fewest sites, with a maximum of 37 occurrences on a single page.

\textbf{False Positive} Certain keywords were more prone to false positives. In our dataset, 60 webpages were incorrectly flagged due to the keyword ``bet'', as substring matching in dorking queries also captured tokens such as ``between'' and ``beta''. In contrast, the keywords ``gacor'', ``maxwin'', and ``togel'' did not produce observable false positives.

\subsection{Defacement Technique}

In our observations, we identified 84 instances in which attackers modified webpage titles. However, we did not observe any defacements that explicitly displayed a “hacked by” message, as reported in prior studies \cite{nguyenDetectingWebsiteDefacement2021}. A closer examination of the defaced webpages revealed that not all defacements involved direct visual alterations, such as modifications to titles, embedded images, or visible text. In several cases, the webpages themselves were not directly altered but instead configured to redirect visitors to external sites.

\textbf{Obfuscation} The attackers behind website defacements containing IOG promotions or links do not always display their work openly. In more than a hundred defaced websites, there were no visible signs of alteration.
Upon closer inspection, we found that some attackers obfuscated the defacement content. One common technique involved using the HTML attribute \texttt{style=``display:none''}, which instructs the browser to load and render the element in the DOM but omit it from the visual layout, making it invisible to users while remaining accessible to automated crawlers. This method was observed in 51.7\% webpages. Another technique placed links and keywords far outside the visible viewport, using extreme CSS position values such as left:-50000000px or top:-9999px, making them undetectable on-screen. This off-screen injection method was identified in 22.8\% webpages. Example of the methods can be seen on Lst.\ref{lst:hiddenHTML}.

\begin{figure}
    \centering
    \begin{lstlisting}[caption={Example of hidden HTML code injected into the webpages: Lines 1–4 show an instance of IOG content embedded within hidden HTML elements, while lines 6–8 illustrate IOG content positioned outside the visible screen area.},label={lst:hiddenHTML},language=HTML]
<div style="display: none;">
    <a href="https://IOG.com/">Hidden</a>
    ...
</div>

<div style="position: absolute; left: -9999rem;">
    <a href="https://IOG.com/">Out of screen</a>
    ...
</div>
\end{lstlisting}
\end{figure}

\textbf{Redirection} We observed that attackers employed URL redirection, where users were automatically forwarded to another site after loading compromised page~\cite{wuCloakingRedirectionPreliminary2005}, on several webpages. Redirections were observed on 13 websites (28 webpages). Among these, seven websites redirected to a single IOG-related domain, one website redirected to two domains, three websites redirected to three domains, one website redirected to five domains, and another website redirected to six domains. On stisalmanar.ac.id, where the redirections pointed to websites containing the string ``banci''(meaning ``coward'' in English), suggesting a single operator. On jeep-indonesia.id, three pages (home, harga, contact) were defaced, and each time the redirection target changed, all pages pointed to the same destination.

\subsection{Reaction times} As shown in Tab.\ref{tab:crawlerOverview}, 346 defaced pages were identified during the one-month data collection period. Of these, 287 were flagged as either fixed or repeat defacement (Tab.\ref{sec:methods_flag}), indicating that at least once during the observation period the pages no longer contained IOG-related keywords, either due to repair or temporary inaccessibility to the crawler. Within this group, 102 pages were defaced only once, while the remainder experienced multiple defacement–fix cycles, with the maximum reaching 112 repetitions.

\begin{figure*}
    \centering
    \includegraphics[width=0.9\linewidth]{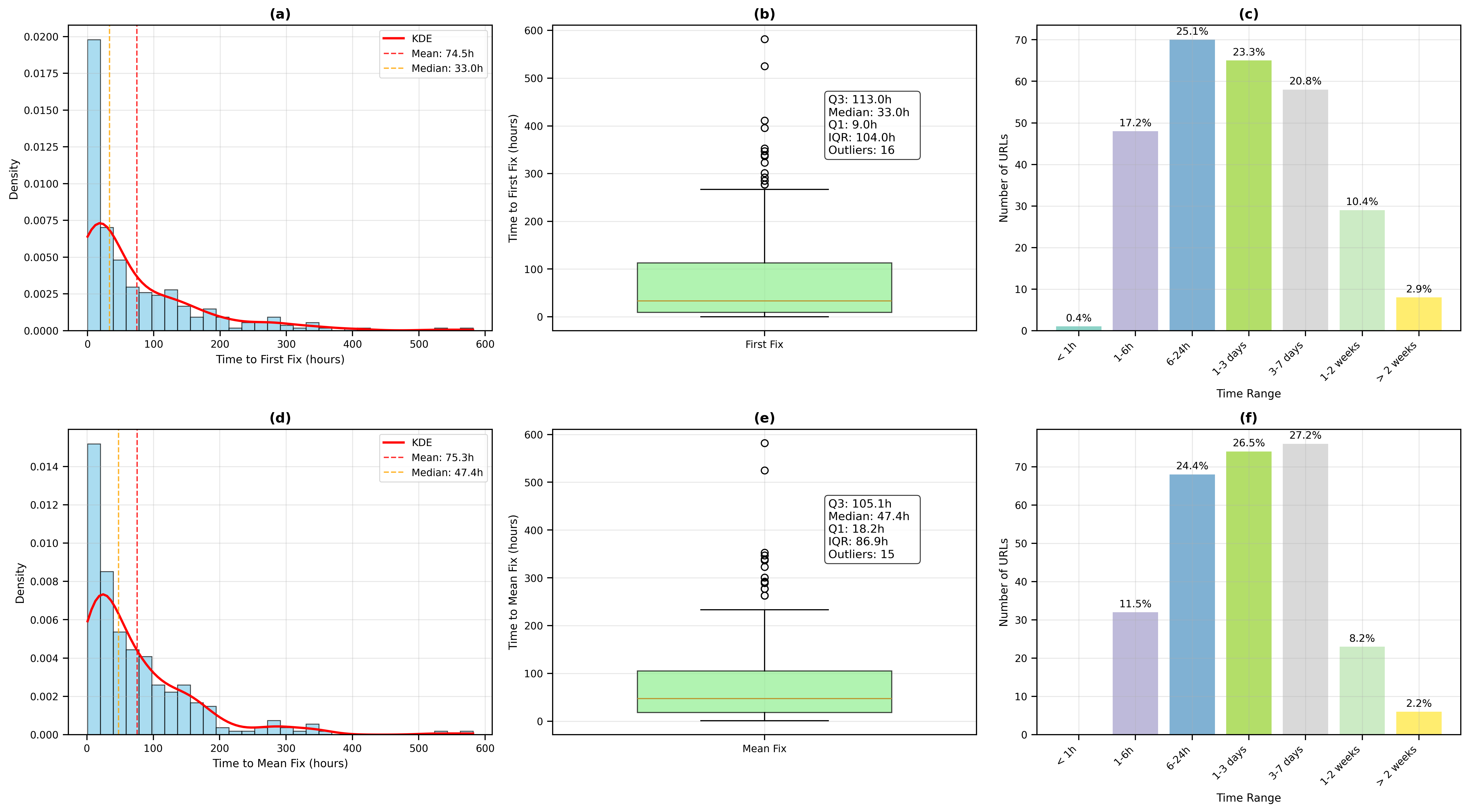}
    \caption{Distribution of $\Delta$ (elapsed time in hours from the first positive keyword detection to the first zero-keyword occurrence): (a) histogram of $\Delta_{first}$; (b) box plot of $\Delta_{first}$ showing quartiles, median, and outliers; (c) categorical distribution of $\Delta_{first}$ grouped by time ranges; (d) histogram of $\Delta_{avg}$; (e) box plot of $\Delta_{avg}$ showing quartiles, median, and outliers; and (f) categorical distribution of $\Delta_{avg}$ grouped by time ranges.}
    \label{fig:time_distribution}
\end{figure*}

For analysis, we distinguish between two measures of remediation. The first is the initial reaction time ($\Delta_{first}$), defined as the time taken to remediate a defacement for the first time (Fig.~\ref{fig:time_distribution}). The second is the average reaction time ($\Delta_{avg}$), defined as the mean remediation time across all cycles for the same page. The mean initial reaction time was 74.7 hours (median 32.7 hours), while the mean average reaction time was 75.3 hours (median 47.4 hours). These values are comparable to Bartoli et al.~\cite{bartoliReactionTimeWeb2009}, who reported an average remediation time of 72.4 hours. Although the mean values are close, the higher median of $\Delta_{avg}$ suggests that websites experiencing repeated defacements tended to require longer recovery in subsequent incidents, possibly due to attacker persistence or slower administrative response after compromise.

In terms of remediation, 121 of 279 URLs (43.4\%) were fixed within 24 hours of their first defacement, compared to 101 of 279 (36.2\%) when considering average recovery time across cycles. Conversely, 37 URLs (13.3\%) required more than a week to recover from their initial defacement, while only 29 (10.4\%) exceeded a week on average. These results suggest that administrators often acted promptly after the first incident, but repeated defacement cycles introduced variability, leading to slower median recovery yet fewer prolonged cases.

\subsection{Internal link} We analyzed the occurrence of third-party URLs on defaced webpages and identified 8,837 unique URLs associated with 5,930 distinct domains. Among these, 3,165 URLs appeared only once, while the mean and median occurrences across webpages were 3.8 and 3, respectively. At the other extreme, nine third-party URLs appeared on 21 different defaced webpages. The three most frequently observed third-party domains were \textit{heylink.me} (present in 16 domains), \textit{lazada.co.id} (9), and \textit{linklst.bio} (9). Both \textit{heylink.me} and \textit{linklst.bio} operate as link-shortening services, suggesting that attackers may have used them to manage and conceal redirection targets. In contrast, \textit{lazada.co.id} is a well-known e-commerce platform, and its appearance on defaced webpages is unexpected, as no assets or content delivery network (CDN) resources from Lazada were used. The domain appeared in nine distinct URL forms, four of which contained an identical SCM query parameter, which may indicate an analytics-related function.

\section{Future Works}

Future study should extend the observation period beyond the one-month window used in this study to capture longer-term patterns of defacement and remediation. In addition, while this study monitored internal links holistically, future work should examine them in greater detail, as not all links redirected to gambling sites, suggesting selective manipulation by attackers. Another direction is to analyze the underlying technology stacks of defaced websites, which may reveal correlations between specific frameworks, servers, or content management systems and their susceptibility to defacement.

\section{Conclusions}
Our study shows that defacement of Indonesian websites remains an active and evolving threat. In just one month of applying keyword-driven dorking combined with systematic crawling, we identified 453 webpages containing illegal online gambling content. While dorking alone is prone to errors, with a false positive rate of about 20.3\% depending on the keywords, integrating crawling and keyword-counting allowed us to separate false from true positives in a lightweight and efficient manner.

The measurements highlight the diversity of defacement techniques. Beyond direct content changes, attackers often used redirection or hidden URLs. Among the cases, 150 webpages experienced repeat defacement, 129 were later fixed, 55 showed keyword modifications, and only 12 remained unchanged. We also observed 84 altered titles, though none included explicit ``hacked by'' signatures. Additionally, 8,837 unique third-party URLs (from 5,930 domains) were injected into defaced pages, most appearing only once, while a small set recurred across multiple sites.

Website responses varied: some domains took no action (67 pages), possible unaware of the defacement, while others attempted repairs. The mean initial reaction time was 74.7 hours, with an overall average of 75.3 hours, indicating inconsistent and often delayed remediation

Overall, our findings demonstrate that simple yet systematic techniques such as dorking and crawling can provide valuable insights into the scale, persistence, and dynamics of defacement, even when compared to more complex machine learning approaches. This paper underscores the need for continuous measurement of defacement activity, not only to support more effective detection strategies but also to provide actionable data that can help institutions improve their incident response practices and strengthen defenses against the growing threat of illicit online gambling.

\section*{Code and Dataset}
After the paper’s acceptance, our code and most of the dataset will be made publicly available.


\bibliographystyle{ieeetr}
\bibliography{ref}

\end{document}